# A Security Protocol for the Identification and Data Encrypt Key Management of Secure Mobile Devices

Chol-Un Kim [1], Dok-Jun An [1, 2] and Song Han [1]

[1] *Faculty of Mathematics, **Kim Il Sung** University, Pyongyang, D.P.R.K.*

[2] Corresponding author e-mail address: myongcholri@yahoo.com

*Abstract* : In this paper, we proposed an identification and data encrypt key manage protocol that can be used in some security system based on such secure devices as secure USB memories or RFIDs, which are widely used for identifying persons or other objects recently. In general, the default functions of the security system using a mobile device are the authentication for the owner of the device and secure storage of data stored on the device.

We proposed a security model that consists of the server and mobile devices in order to realize these security features. In this model we defined the secure communication protocol for the authentication and management of data encryption keys using a private key encryption algorithm with the public key between the server and mobile devices. In addition, we was performed the analysis for the attack to the communication protocol between the mobile device and server.

Using the communication protocol, the system will attempt to authenticate the mobile device. The data decrypt key is transmitted only if the authentication process is successful. The data in the mobile device can be decrypted using the key.

Our analysis proved that this Protocol ensures anonymity, prevents replay attacks and realizes the interactive identification between the security devices and the authentication server.

# 1. Introduction

Massive use of mobile storage media raises data security and user authentication problem seriously.

Reliability of security system should not be based on system mechanism or complexity of system analysis, and it should guarantee safety even if system mechanism or encryption algorithm is opened to third party.

In the paper, we proposed an identification and data encrypt key manage protocol for using in some user authentication and dik encrypt system based on such secure devices as secure USB memories or RFIDs.

We analyze security problems of existing disk encryption methods and identification protocol in section 2 and describe our identification protocol and data encrypt key manage protocol, in section 3. In section 4, we analyze some security performance of our system.

# 2. Previous Works

Nowadays, there are a lot of security system that are using such storage devices being capable of storing personal data securely as secure USB, smart card or RFIDs.

Several different methods for disk encryption, such as LoopAES, EFS, TrueCrypt, NCryptfs, were suggested [ 5, 6, 8, 9, 10].

In these methods, encryption of disk block is expressed as follows.

$$C=OP(BE(OP(P, DEK, i), DEK), DEK, i) \qquad (1)$$

Here, BE is a block encryption function (AES, 3DES, etc.), OP is an operation function (CBC, LRW, XTS, etc.), DEK is a disk encryption key, P is a plaintext, i is a block index and C is a cipher text.

As expression (1) shows, these disk encryption systems encrypt plaintext using symmetric-key algorithm through certain operation and apply this operation again to encrypted result.

These disk encryption methods have some weakness in terms of time passage and space expansion.
- Temporal limitation
    If third party succeeds to detect encryption key of a certain sector, data which is stored in this sector later can be decoded.
- Spatial limitation
    If third party succeeds to detect encryption key of a certain block, whole data of disk is in danger of being decoded.

GBDE based encryption method which was implemented in FreeBSD overcame these temporal and spatial limitations to a certain extent [7].

Although GBDE based disk encryption method overcomes temporal and spatial limitations of previous disk encryption methods considerably, it still has some security problems to be solved.
- Key-key for a given sector is fixed because it is decided depending on the sector address. That is, when it was written new data on sector, sector key is encrypted by same key-key.
- It is easy to get keychain used to encrypt plaintext data, if the correlation between random data generated consecutively by PRNG is revealed, because it directly uses random data generated by PRNG as key for plaintext.

the SDMS Encrypt Method in [1] solved temporal limitation problem and spatial limitation problem of data encryption to a certain extent.

And, SDMS can control security performance flexibly according to the security requirements.

The master data encrypt key DEK in SDMS must be encrypted based on the user authentication information. However, it did not propose a specific Authentication protocol.

There are also a lot of authentication protocols that are using mobile devices to identify the identity of their users [2-4]. But there are many protocols that have various volunerabilities or not to be satisfied the requirements for applications.

For example, the protocols in [2, 3] expose the device's ID, or cannot to be clear the requirements of anonymity and resistence for device traceability because of using fixed hash value for each device, and the protocols in [4] have the problem that the load of authentication servers are rapidly increasing as the number of devices grows.

In the next section, we present a user identification and data encrypt key manage protocol.

### 3. Our Identification and Data Encrypt Key Management Protocol

In this paper, the authentication system consists of the **secure storage devices** being capable of storing the legitimate user's ID and having some cryptographic computation capability such as secure USB memories, smart cards or RFIDs, the **authentication server** that maintains the IDs of all legitimate users on their internal database and confirms the existence of the specific user's ID in the ID database responding to authentication requests, and the **brokers** which are placed between the secure devices and authentication server and transmit any messages of them (for example, the use authentication software on the computer that the secure devices are connected).

#### 3.1 Notations

$D$:  A specific (security) device

$ID_D$:  The ID of the device $D$ (namely, the ID of the legistimate device owner that the system has to identify)

$r_D$:  The random number generated by the authentication server and

stored in the device $D$.

*Hello* : The ID query message that the broker send to the secure storage device.

$H : \{0, 1\}^* \to \{0, 1\}^h$ : The h-bit cryptographic hash function that have the preimage resistence and the collision resistence cryptographic hash function ($h$ is a positive integer)

*PRNG* : The pseudo-random generator implemented in the authentication server. We suppose that the bit length of random number generated by it cannot be beyond the one of hash value of $H$ ($h$).

$PKE = (KeyGen^P, E^P, D^P)$ : The secure public-key cryptosystems implemented in the authentication server and secure storage devices.

$(e^P, d^P)$ : The pair of public and private key of $PKE$, generated by $KeyGen^P$. It assumes that the decryption key (the private key) is only known to the authentication server.

$SKE = (KeyGen^S, E^S, D^S)$ : The secure symmetric-key cryptosystems implemented in the authentication server and secure storage devices.

PBKDF2 ($P$, $S$, $c$, $dkLen$): The key derive function defined in the PKCD#5 v2.0 [11].

## 3.2 The Structure of Secure Storage Device's Signature

This secure storage device's signature $Sig_D$ consists of $ID_D$ and key for the secure storage devices $D$.

Here, $ID_D$ is the unique random number for the device $D$, and key is the encrypt/decrypt key for data in the device $D$.

All signature $Sig_D$ for any secure storage devices $D$ are stored in the authentication server.

The structure of signature $Sig_D$ in the server is as follows:

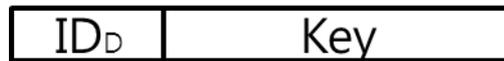

Figure. Structure of signature $Sig_D$

## 3.3 Our Identification and Key Management Protocol

Our protocol consists of the preparation step and the authentication step.

**[ The Preparation Step]**

Any secure storage device registered in the authentication system, $D$ are maintained the random number $r_D$ generated by the authentication server. The random number $r_D$ is unique for

Security protocol for the identification and data encrypt key management of secure mobile devices the device $D$, so it differs by devices. The authentication server must maintain the random numbers in some period for resisting the replay attacks and guarantee their uniqueness. It stores all signature $Sig_D$ for any secure storage devices $D$ in the system to its secure internal database by making $H(ID_D)$ as the search primary key.

**[The Authentication Step]**

**Step 1**: The broker initiats the protocol run by sending the message *Hello* to the device $D$.

**Step 2**: The device $D$ run the following actions by using its ID, $ID_D$ and the random number $r_D$ in its memory..

(2-1) $x \leftarrow H(ID_D) \oplus r_D$

(2-2) $a \leftarrow E^P(e^P, x)$

(2-3) $C \leftarrow E^S(x, r_D)$

(2-4) Send $(a, H(x), C, H(r_D))$ to the authentication server through the broker.

**Step 3**: The authentication server performs the following actions after receiving $(a, H(x), C, H(r_D))$.

(3-1) $x \leftarrow D^P(d^P, a)$

(3-2) $r_D \leftarrow D^S(x, C)$

(3-3) Use $H(x)$ and $H(r_D)$ to verify the accuracy of the $x$ and $r_D$.

(3-4) $H(ID_D) \leftarrow x \oplus r_D$

(3-5) $Sig_D$ derive the value of magnetic material at the base of the search key $H(ID_D)$.

Server will get identification of Security devices after the end of this step.

**Step 4**: The authentication server performs the following behavior.

(4-1) Generate new random number $r_{new}$ that is different $r_D$ used by *PRNG*.

(4-2) $y \leftarrow H(ID_D) \oplus r_{new}$

(4-3) $b \leftarrow E^S(ID_D, r_{new})$

(4-4) $k \leftarrow \text{PBKDF2}(ID_D, r_{new}, c, dkLen)$

Here, $c$ is iteration count, a positive integer and *dkLen* is intended length in octets of the derived key, a length of data encrypt key in Mobile device.

(4-5) $g \leftarrow E^S(k, Sig_D)$

(4-6) Send $(b, H(y), H(r_{new}), g)$ to security device $D$ through the broker.

**Step 5**: The security device $D$ performs the following actions after receiving $(b, H(y), H(r_{new}), g)$

(5-1) $r_{new} \leftarrow D^S(ID_D, b)$

(5-2) Confirm the accuracy of the $r_{new}$ used by the hash value $H(r_{new})$

(5-3) Compare $H(H(ID_D) \oplus r_{new})$ and the receipted hash value $H(y)$

If the two values match, the security device will get the identification and reliability for $r_{new}$, based on the (5-1).

(5-4)  $k \leftarrow \text{PBKDF2}(ID_D, r_{new}, c, \text{dkLen})$

(5-5)  $Sig \leftarrow D^S(k, g)$

(5-6) Compare $ID_D$ of $Sig$ and the $ID_D$ of the security device.
  If the two values match, $r_{new}$ should be kept and data in the security device should be decrypt using key of $Sig$.

## 4. Security Interpretations

[**Theorem 1**] *This Protocol ensures anonymity*.

(**Proof**) Message that is sent Security devices to the server, are included $(a, H(x), C, H(r_D))$.

If you guarantee safety for $PKE$, the attacker can't get accurate $x$, and because don't know the encryption key that was used to encrypt the symmetric key, Ciphertext for decryption is impossible, therefore don't know $r_D$.

The attacker never knows $H(ID_D)$. (End)

[**Theorem 2**] *These Protocol conventions prevent replay attacks*.

(**Proof**) In the Protocol tags and serve are used temporary as a random number $r_D$ only for the duration of the conversation and it will be updated after successful authentication.

Also because All Information of Conventions notice depend random number $r_D$, replay attacks is impossible. (End)

[**Theorem 3**] *This Protocol realizes the interactive identification between the security devices and the authentication server*.

(**Proof**) Because identify a number $ID_D$ of physical security devices is known only self-security devices and authentication Server, in step3 (3-5) Server will confirm the identity of the security device, and in step 5(5-3) security device will confirm the identity of the server. (End)

## 5. Conclusions

1) The covenant presented in this paper are able to be used in one computer system security, remote user authentication and  Data Encrypt Key management, which are based on software security USB, RFID and etc. And, the protocol can be used to build a security system within the company through a mobile device such as USB.

2) The mobile devices does not contain any information that can help to get the data encryption key, so the data in the device is secure even if the attacker illegally owned mobile devices.

**Acknowledgement:** Authors would like to thank anonymous reviewers' help and advice.

**References**

[1] **Myongchol RI, Dokjun An and Changil Choe**: Secure Disk Mixed System. Journal of Mobile, Embedded and Distributed Systems, Vol. 4, No. 4, 2012. Cross-ref; arXiv:1212.2054[cs.CR]

[2] **S.A. Weis, S.E. Sarma, R.L. Rivest, D.W. Engels**: Security & Privacy Aspects of Low-Cost Radio Frequency Identification Systems, Security in Pervasive Computing 2003, LNCS 2802, 201~212 (2004)

[3] **A.D. Henrici, P. Mauller**: Hash-based enhancement of location privacy for radio-frequency identification devices using varying identifiers, PerSec'04 at IEEE PerCom, 149~153 (2004)

[4] **J.C. Chang, H.L. Wu**: A Hybrid RFID Protocol against Tracking Attacks, IIH-MSP'2009.1 12~14(2009)

[5] **Tom Olzak**: Evaluation of TrueCrypt as a Mobile Data Encryption Solution.2008. URL

[6] **Clemens Fruhwirth**: Hard disk encryption with DM-Crypt, LUKS, and cryptsetup. ISSUE 61, 2005. 65-71. URL

[7] **P.H.Kamp**: GBDE-GEOM Based Disk Encryption. BSD Con'03, 57-68, 2003.

[8] **Jari Ruusu**: LoopAES, 2005, URL http://loop-aes.sourceforge.net/

[9] **Microsoft Corporation**: Encrypting File System for Windows 2000. Technical report, 1999.

[10] **S. Shepler**: NFS Version 4 Protocol. Technical Report RFC 3010, Network Working Group, 2000.

[11] **RSA Laboratories**: Password-Based Cryptography Standard. Technical Report PKCS #5, RSA Data Security,1999.

[12] **W. Stallings**: Cryptography and Network Security. Principles and Practices. Fourth Edition, Prentice Hall, 592 pages, 2006.

[13] **Rick Lehtinen**: Computer Security Basics. 2nd Edition, O'Reilly, 310pages, 2006.